# Integrated flat-top reflection filters operating near bound states in the continuum

LEONID L. DOSKOLOVICH,[1,2,3] EVGENI A. BEZUS,[1,2,3,*] AND DMITRY A. BYKOV[1,2]

[1]*Image Processing Systems Institute — Branch of the Federal Scientific Research Centre "Crystallography and Photonics" of Russian Academy of Sciences, 151 Molodogvardeyskaya st., Samara 443001, Russia*
[2]*Samara National Research University, 34 Moskovskoye shosse, Samara 443086, Russia*
[3]*Contributed equally to this work*
*\*evgeni.bezus@gmail.com*

**Abstract:** We propose and theoretically and numerically investigate narrowband integrated filters consisting of identical resonant dielectric ridges on the surface of a single-mode dielectric slab waveguide. The proposed composite structures operate near a bound state in the continuum (BIC) and enable spectral filtering of transverse-electric-polarized guided modes propagating in the waveguide. We demonstrate that by proper choice of the distances between the ridges, flat-top reflectance profiles with steep slopes and virtually no sidelobes can be obtained using just a few ridges. In particular, the structure consisting of two ridges can optically implement the second-order Butterworth filter, whereas at a larger number of ridges, excellent approximations to higher-order Butterworth filters can be achieved. Owing to the BIC supported by the ridges constituting the composite structure, the flat-top reflection band can be made arbitrarily narrow without increasing the structure size. In addition to the filtering properties, the investigated structures support another type of BICs — Fabry–Pérot BICs arising when the distances between the adjacent ridges meet the Fabry–Pérot resonance condition. In the vicinity of the Fabry–Pérot BICs, an effect similar to the electromagnetically induced transparency is observed, namely, sharp transmittance peaks against the background of a wide transmittance dip.

## 1. Introduction

Spectral filters that selectively reflect or transmit the incident light are indispensable in various optical devices including spectrum analyzers, image sensors and wavelength demultiplexers. In a wide class of planar (integrated) optoelectronic systems, spectral or spatial filtering is performed in a slab waveguide [1–7]. Such geometry corresponds to the "insulator-on-insulator" platform and is suitable for the creation of fully integrated optical devices.

In [1, 8–12], planar Bragg gratings (BGs) and phase-shifted Bragg gratings (PSBGs) for spectral filtering of optical radiation propagating in the waveguide were proposed. A planar BG corresponds to a single-mode slab waveguide with a periodically corrugated surface and acts as a reflection filter centered at the Bragg wavelength. A change in the waveguide thickness leads to a change in the effective refractive index of the guided mode, which makes it possible to encode the required refractive index distribution analogous to a free-space BG. A planar PSBG consists of two or more symmetric BGs separated by phase-shift regions and enables obtaining a narrow transmittance peak at the center of the stopband [1, 11, 12].

Recently, a promising alternative to a narrowband planar BG-based filter was proposed [13–16]. It was shown that an extremely simple resonant structure consisting of a single dielectric ridge on the surface of a single-mode slab waveguide supports high-Q resonances associated with the excitation of the TM-polarized modes of the ridge by the obliquely incident TE-polarized mode of the waveguide. Moreover, at properly chosen parameters, the ridge supports bound states in the continuum (BICs). Near a BIC, the reflectance spectrum has a symmetric Lorentzian line shape, which makes it possible to use the ridge structure as a narrowband filter operating in reflection. At the same time, the Lorentzian line shape of the reflectance peak does not always fit the design requirements. For many practical applications, achieving nearly rectangular reflectance spectra with flat top and steep slopes is highly desirable [11, 17].

In the present work, we investigate composite resonant structures comprising several identical ridges on the surface of a slab waveguide separated by phase-shift regions. We show that by proper choice of the widths of the phase-shift regions, flat-top reflectance profiles with steep slopes can be easily obtained using just a few ridges. Moreover, the width of the reflection band near the BICs can be arbitrarily small by choosing the ridge width. To the best of our knowledge, this work presents the first theoretical and numerical demonstration of integrated flat-top filters operating in the near-BIC regime. In addition to the applications in spectral filtering, the proposed composite structures have another interesting property. At the distances between the ridges, which satisfy the condition of the Fabry–Pérot resonance, $(N-1)$ - degenerate BICs of Fabry–Pérot type arise in the composite structure comprising $N$ ridges. In the vicinity of these BICs, an effect similar to the electromagnetically induced transparency is observed.

## 2. Geometry and reflectance spectra of a single ridge on the surface of a slab waveguide

One of the simplest integrated structures possessing remarkable resonant properties is a dielectric ridge located on the surface of a single-mode slab waveguide [Fig. 1(a)]. In our recent works [13, 14], it is shown that in the case of diffraction of an obliquely incident TE-polarized mode of the waveguide, the ridge exhibits BICs and high-Q resonances associated with the excitation of the cross-polarized modes of the ridge.

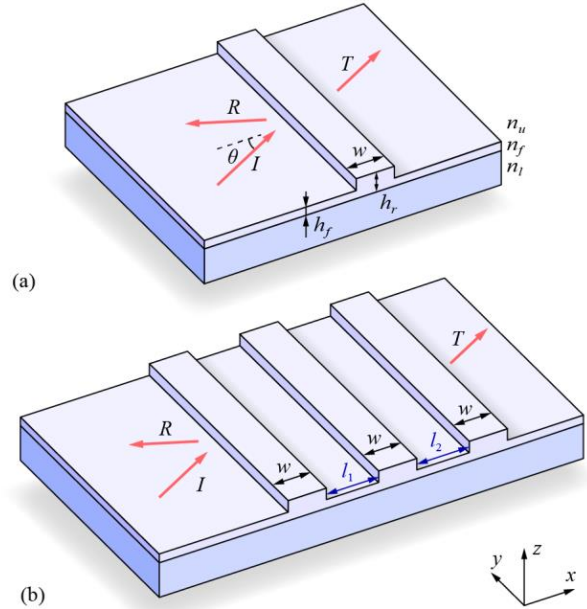

Fig. 1. Geometry of a ridge on a waveguide layer (a) and of a composite structure consisting of three ridges separated by phase-shift regions (b). The red arrows indicate the propagation directions of the incident wave *I*, reflected wave *R* and transmitted wave *T*.

As an example, Fig. 2 shows the TE-polarized mode reflectance vs. the ridge width *w* and the angle of incidence $\theta$. The plot was calculated using an efficient in-house implementation of the aperiodic rigorous coupled-wave analysis (aRCWA) technique [18–20]. The simulations were carried out with the following parameters: free-space wavelength $\lambda_0 = 630\,\text{nm}$, refractive index of the waveguide core layer $n_f = 3.3212$ (GaP), refractive indices of the superstrate and the substrate $n_u = 1$ and $n_l = 1.45$ (fused silica), respectively; the thickness of the waveguide core layer $h_f = 80\,\text{nm}$, and the thickness of the waveguide in the ridge region $h_r = 110\,\text{nm}$. At these parameters, the waveguide is single-mode both outside and inside the ridge region. Effective refractive indices of the TE- and TM-polarized modes supported by the waveguide amount to $n_{wg,TE} = 2.5913$ and $n_{wg,TM} = 1.6327$, respectively. In the ridge region (i.e. at the waveguide thickness $h_r = 110\,\text{nm}$), effective refractive indices equal $n_{r,TE} = 2.8192$ and $n_{r,TM} = 2.1867$.

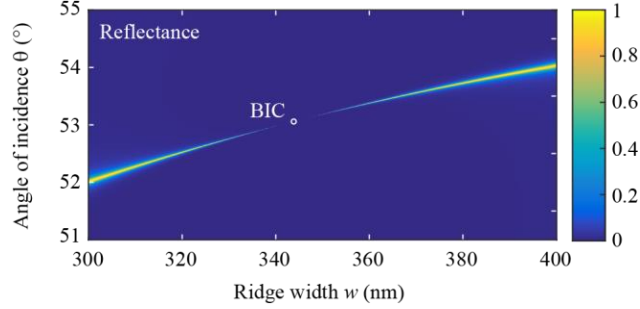

Fig. 2. Reflectance of an obliquely incident TE-polarized guided mode from the ridge vs. the ridge width $w$ and the angle of incidence $\theta$. The white circle indicates the BIC position.

The reflectance spectrum shown in Fig. 2 has a pronounced resonant maximum. As it was shown in [13], this resonance is caused by the excitation of a cross-polarized ("TM-like") mode of the ridge. It is important to note that in the chosen range of angles of incidence $\theta \in [51°, 55°]$, no out-of-plane scattering of the incident mode on the ridge occurs. Moreover, no cross-polarized reflected and transmitted modes are excited. A detailed description of this scattering cancellation mechanism is presented in [13, 14]. In this case, at maxima, the reflectance reaches unity. The angular width of the resonant maximum varies from approximately $0.1°$ at the boundaries of the considered interval of the ridge width $w \in [300, 400]$ nm to zero at $w = 344$ nm. The point, at which the resonance disappears ($w = 344$ nm, $\theta = 53.06°$), corresponds to a BIC. Angular and wavelength reflectance spectra in the vicinity of a BIC have Lorentzian line shapes [13, 14]. As an example, Fig. 3 shows the reflectance spectra at $w = 355$ nm, $w = 360$ nm and $w = 380$ nm. The corresponding angular spectra are shown in Fig. 3(a). The maximum of the reflectance is reached at the angles of incidence $\theta_1 = 53.28°$ ($w = 355$ nm), $\theta_2 = 53.37°$ ($w = 360$ nm), and $\theta_3 = 53.72°$ ($w = 380$ nm). Figure 3(b) shows the wavelength spectra calculated at the corresponding angles of incidence $\theta_i$, $i = 1, 2, 3$. It is evident from Fig. 3 that the width of the reflectance peak decreases when approaching the BIC. Theoretically, in the vicinity of a BIC, it is possible to obtain a reflectance peak having an arbitrarily small width.

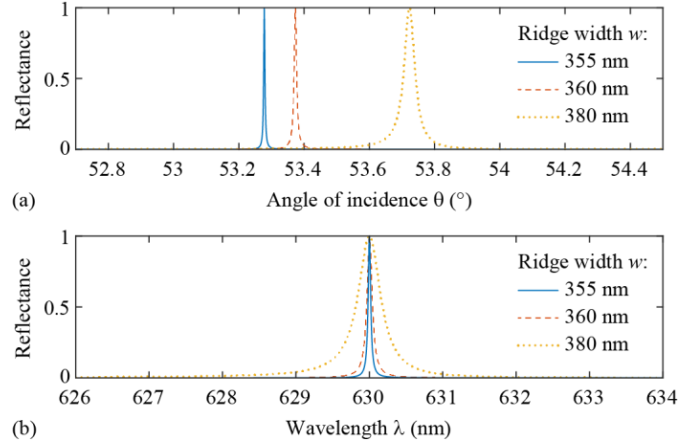

Fig. 3. Angular (a) and wavelength (b) ridge reflectance spectra at $w = 355$ nm (solid blue lines), $w = 360$ nm (dashed red lines) and $w = 380$ nm (dotted yellow lines).

## 3. Theoretical model describing the spectra of a composite structure comprising several ridges

Lorentzian shape of the reflectance spectra shown in Fig. 3 makes it possible to use the ridge as an angular or as a frequency filter. However, a rectangular reflectance peak with flat top, steep slopes and low sidebands would be more suitable for many practical applications [11, 17]. In what follows, we demonstrate that composite structures comprising several ridges on the surface of a slab waveguide, which operate in the near-BIC regime, enable obtaining narrow flat-top reflectance peaks with steep slopes. In this case, the BICs supported by the ridges allow one to make the width of the flat-top reflection band arbitrarily small by choosing the ridge width.

Geometry of the proposed composite structure is shown in Fig. 1(b). For illustrative purposes, the structure consisting of three ridges is depicted. We assume that the ridges constituting the composite structure are identical, whereas the widths of the phase-shift regions separating the ridges may, in general, be different.

To analyze the composite structures, it is convenient to describe the optical properties of a single ridge at a fixed angle of incidence $\theta$ by the scattering matrix [20–22]

$$\mathbf{S}_1(\lambda) = \begin{pmatrix} t_1(\lambda) & r_1(\lambda) \\ r_1(\lambda) & t_1(\lambda) \end{pmatrix}, \quad (1)$$

where $r_1(\lambda)$ and $t_1(\lambda)$ are the complex reflection and transmission coefficients, respectively, which are considered as functions of the wavelength. For Lorentzian resonances (see Fig. 3), these coefficients can be approximated by the following expressions:

$$r_1(\lambda) = \exp\{i\varphi\} \frac{i \operatorname{Im} \lambda_p}{\lambda - \lambda_p},$$
$$t_1(\lambda) = \exp\{i\varphi\} \frac{\lambda - \operatorname{Re} \lambda_p}{\lambda - \lambda_p}, \quad (2)$$

where the pole $\lambda_p$ is the complex wavelength of the eigenmode supported by the ridge. This form of the reflection and transmission coefficients ensures the unitarity of the scattering matrix of Eq. (1). According to Eq. (2), the reflectance $R_1(\lambda) = |r_1(\lambda)|^2 = 1 - |t_1(\lambda)|^2$ reaches unity at

$\lambda = \lambda_0 = \operatorname{Re}\lambda_p$. The width of the resonance is determined by the imaginary part of the pole $\lambda_p$. Indeed, from Eq. (2), it is easy to obtain that the full width at half maximum of the resonant reflectance peak $R(\lambda)$ (and of the resonant transmittance dip $T(\lambda) = |t_1(\lambda)|^2$) amounts to $\Delta = 2|\operatorname{Im}\lambda_p|$. Note that Eq. (2) remains valid in the vicinity of a BIC. If the BIC condition is strictly satisfied (i.e. $\operatorname{Im}\lambda_p = 0$), the resonance disappears (see Fig. 2), the reflection coefficient $r_1(\lambda)$ vanishes, and the magnitude of the transmission coefficient becomes unity: $|t_1(\lambda)| = 1$.

## 3.1 Two-ridge composite structure

Let us now consider a composite structure consisting of two ridges described by identical scattering matrices defined by Eq. (1) and separated by a phase shift region with the width $l_1$. At a fixed angle of incidence $\theta_0$, the scattering matrix of the composite structure can be expressed through the matrix $\mathbf{S}_1(\lambda)$ as [20, 23]

$$\mathbf{S}_2(\lambda) = \mathbf{S}_1(\lambda) * \mathbf{L}(l_1) * \mathbf{S}_1(\lambda), \tag{3}$$

where the symbol $*$ denotes the Redheffer star product [20], and $\mathbf{L}(l_1)$ is the scattering matrix describing the phase shift region:

$$\mathbf{L}(l_1) = \exp\{i\psi(l_1)\}\mathbf{I}. \tag{4}$$

Here, $\psi(l_1)$ is the phase shift acquired by the mode upon propagation through the phase-shift region, and $\mathbf{I}$ is the $2\times 2$ identity matrix. For simplicity, in Eq. (4) we assume that the dependence of the phase shift on the wavelength can be neglected, so that this phase shift can be approximated as

$$\psi(l_1) = k_x l_1 = k_0 n_{wg,TE} \cos\theta_0 \cdot l_1, \tag{5}$$

where $k_x = k_0 n_{wg,TE} \cos\theta_0$ is the $x$-component of the wave vector of the incident TE-polarized guided mode, $k_0 = 2\pi/\lambda_0$ is the wavenumber, and $n_{wg,TE}$ is the effective refractive index of the incident TE-polarized mode. Note that the made assumption is valid when we consider a narrow wavelength range centered at the resonant wavelength $\lambda_0 = \operatorname{Re}\lambda_p$.

Using the definition of the Redheffer star product [20], one can easily obtain the scattering matrix of the composite structure in the following form:

$$\mathbf{S}_2(\lambda) = \begin{pmatrix} t_2(\lambda) & r_2(\lambda) \\ r_2(\lambda) & t_2(\lambda) \end{pmatrix} = \frac{1}{1-\exp\{2i\psi\}r_1^2} \begin{pmatrix} \exp\{2i\psi\}t_1^2 & r_1\left[1-\exp\{2i\psi\}\left(r_1^2-t_1^2\right)\right] \\ r_1\left[1-\exp\{2i\psi\}\left(r_1^2-t_1^2\right)\right] & \exp\{2i\psi\}t_1^2 \end{pmatrix}. \tag{6}$$

By substituting Eq. (2) into Eq. (6), we obtain the following resonant approximations of the reflection and transmission coefficients of the composite structure:

$$r_2(\lambda) = \gamma_{r,2}\frac{\lambda-\lambda_{z,1}}{(\lambda-\lambda_{p,1})(\lambda-\lambda_{p,2})}, \quad t_2(\lambda) = \gamma_{t,2}\frac{(\lambda-\operatorname{Re}\lambda_p)^2}{(\lambda-\lambda_{p,1})(\lambda-\lambda_{p,2})}, \tag{7}$$

where

$$\lambda_{z,1} = \operatorname{Re}\lambda_p + \operatorname{tg}(\psi(l_1)+\varphi)\operatorname{Im}\lambda_p, \tag{8}$$

$$\lambda_{p,1,2} = \operatorname{Re}\lambda_p + i(1\pm\sigma)\operatorname{Im}\lambda_p \tag{9}$$

$\gamma_{t,2} = \exp\{i\psi(l_1)+2i\varphi\}$, $\gamma_{r,2} = 2i\gamma_{t,2}\cos(\psi(l_1)+\varphi)\operatorname{Im}\lambda_p$, and $\sigma = \exp\{i\psi(l_1)+i\varphi\}$.
According to Eqs. (7)–(9), the composite structure consisting of two ridges supports two eigenmodes. These eigenmodes have the complex wavelengths defined by Eq. (9), which are the poles of the reflection and the transmission coefficients. Besides, the reflection coefficient has a real-valued zero $\lambda = \lambda_{z,1}$ defined by Eq. (8), whereas the transmission coefficient has a real-valued second-order zero at $\lambda = \operatorname{Re}\lambda_p$.

It is interesting to discuss the behavior of Eq. (7) in the case when the two ridges constituting the composite structure support a BIC ($\lambda_p \in \mathbb{R}$). In this case, the imaginary part $\operatorname{Im}\lambda_p$ in Eqs. (8) and (9) vanishes, and $\lambda_{p,1} = \lambda_{p,2} = \lambda_{z,1} = \operatorname{Re}\lambda_p = \lambda_p$. Therefore, taking into account the expression for $\gamma_{r,2}$, the two poles in both fractions in Eq. (7) are canceled out by the zeros. Therefore, the considered composite structure supports two BICs having the same wavelength $\lambda_p$, or, in other words, a doubly-degenerate BIC.

It is evident from Eq. (7) that in the general case ($\operatorname{Im}\lambda_p > 0$), the expressions for the reflection and the transmission coefficients $r_2(\lambda)$ and $t_2(\lambda)$ have a significantly more complex form than the corresponding coefficients $r_1(\lambda)$ and $t_1(\lambda)$ of a single ridge. This gives more flexibility for the control of the resonant peak shape. In particular, under the condition

$$\psi(l_1)+\varphi = \pi(m-1/2),\ m\in\mathbb{N}, \tag{10}$$

where the phase shift $\psi(l_1)$ is defined by Eq. (5) and $\varphi$ is the phase of the reflection coefficient $r_1(\lambda_0)$ [Eq. (2)], the coefficients $r_2(\lambda)$ and $t_2(\lambda)$ take the form

$$r_2(\lambda) = \frac{2\exp\{i\varphi\}\operatorname{Im}\lambda_p}{1-[(\lambda-\lambda_p)/\operatorname{Im}\lambda_p]^2},$$
$$t_2(\lambda) = (-1)^m i\exp\{i\varphi\}\frac{(\lambda-\operatorname{Re}\lambda_p)/\operatorname{Im}\lambda_p}{1-[(\lambda-\lambda_p)/\operatorname{Im}\lambda_p]^2}. \tag{11}$$

It is important to note that the expression for $r_2(\lambda)$ in Eq. (11) coincides with the transfer function of the second-order Butterworth filter and, compared to $r_1(\lambda)$, provides a much more rectangular shape of the reflectance peak.

### 3.2 Multiple-ridge composite structures

Similarly, one can obtain the reflection and the transmission coefficients of the composite structure consisting of three equally spaced ridges. Indeed, by calculating $\mathbf{S}_3(\lambda) = \mathbf{S}_2(\lambda)*\mathbf{L}(l_1)*\mathbf{S}_1(\lambda)$, we obtain the reflection and the transmission coefficients of the composite structure in the form

$$r_3(\lambda) = \gamma_{r,3} \frac{(\lambda - \lambda_{z,1})(\lambda - \lambda_{z,2})}{(\lambda - \lambda_{p,1})(\lambda - \lambda_{p,2})(\lambda - \lambda_{p,3})},$$
$$t_3(\lambda) = \gamma_{t,3} \frac{(\lambda - \operatorname{Re}\lambda_p)^3}{(\lambda - \lambda_{p,1})(\lambda - \lambda_{p,2})(\lambda - \lambda_{p,3})},$$
(12)

where

$$\lambda_{z,1,2} = \operatorname{Re}\lambda_p + i\frac{1-\sigma^2}{1\pm\sigma+\sigma^2}\operatorname{Im}\lambda_p,$$
(13)

$$\lambda_{p,1} = \operatorname{Re}\lambda_p + i(1-\sigma^2)\operatorname{Im}\lambda_p,$$
$$\lambda_{p,2,3} = \lambda_p + i\frac{\sigma}{2}\left[\sigma \pm \sqrt{8+\sigma^2}\right]\operatorname{Im}\lambda_p,$$
(14)

where $\sigma = \exp\{i\psi(l_1) + i\varphi\}$, $\gamma_{r,3} = i\exp\{i\varphi\}\left[1 + 2\sigma^3 \cos(\psi(l_1)+\varphi)\right]\operatorname{Im}\lambda_p$, and $\gamma_{t,3} = \exp\{2i\psi(l_1) + 3i\varphi\}$. According to Eqs. (12)–(14), the reflection and the transmission coefficients of the composite structure comprising three ridges have three poles [three eigenmodes with the complex wavelengths defined by Eq. (14)]. The reflection coefficient has two complex-valued zeros defined by Eq. (13), whereas the transmission coefficient has a third-order zero $\lambda = \operatorname{Re}\lambda_p$.

In the general case of a composite structure consisting of $N$ ridges, the reflection and the transmission coefficients of such a composite structure have $N$ poles ($N$ eigenmodes with the complex wavelengths $\lambda_{p,m}$) and can be described by the following expressions:

$$r_N(\lambda) = \gamma_{r,N} \frac{\prod_{m=1}^{N-1}(\lambda - \lambda_{z,m})}{\prod_{m=1}^{N}(\lambda - \lambda_{p,m})},$$
$$t_N(\lambda) = \gamma_{t,N} \frac{(\lambda - \operatorname{Re}\lambda_p)^N}{\prod_{m=1}^{N}(\lambda - \lambda_{p,m})}.$$
(15)

Let us note that the reflection coefficient $r_N(\lambda)$ has $N-1$ complex-valued zeros $\lambda_{z,m}$, whereas the transmission coefficient $t_N(\lambda)$ has a real-valued zero $\lambda = \operatorname{Re}\lambda_p$ of the order $N$. If the ridges constituting the composite structure support a BIC, $\lambda_p$ is real and, as one can show similarly to the two-ridge structure discussed above, $\lambda_{p,m} = \lambda_{z,m} = \lambda_p$. Therefore, all the poles in Eq. (15) get canceled out by the zeros. This means that $N$ independent BICs having the same frequency coexist in the structure comprising $N$ ridges. Such BICs can be referred to as $N$-degenerate. Detuning from the BIC condition lifts the degeneracy, which results in $N$ distinct poles (resonances) and $N-1$ complex-valued zeros $\lambda_{z,m}$ of the reflection coefficient. These zeros and poles of the reflection and the transmission coefficients are functions of the widths $l_i$, $i = 1,...,N$ of the phase-shift regions separating the ridges. In the following section, we will show that optimization with respect to these parameters enables obtaining a nearly rectangular reflectance peak with flat top, steep slopes, and no sidelobes.

Let us also emphasize that the theoretical results obtained in this section are not specific for the considered ridge structures and can be applied to any composite structure comprising Lorentzian-line-shape resonators separated by phase-shift layers.

## 4. Numerical investigation and optimization of composite structures

In Section 3, it was shown that if the condition of Eq. (10) is met, the complex reflection coefficient of the composite structure consisting of two ridges coincides with the transfer function of the second-order Butterworth filter, which provides a more rectangular shape of the resonance than that of a single ridge. Unfortunately, this effect is not preserved at a larger number of ridges $N > 2$. For example, Fig. 4 shows the rigorously calculated transmittance and reflectance spectra of the composite structures consisting of $N = 2$, $N = 4$, and $N = 6$ ridges. The spectra were calculated at the angle of incidence $\theta = 53.72°$, ridge width $w = 380$ nm, and a fixed distance between the ridges $l = 948$ nm satisfying the condition of Eq. (10) at $m = 3$. For comparison, the dashed lines in Fig. 4 show the spectra of a single ridge. At $N = 2$ [Fig. 4(a)], the spectra are with a high accuracy described by the expressions $R_2(\lambda) = |r_2(\lambda)|^2$ and $T_2(\lambda) = 1 - |r_2(\lambda)|^2$, where $r_2(\lambda)$ is the transfer function of the second-order Butterworth filter defined by Eq. (11). With an increase in $N$, the central parts (reflectance peak and transmittance dip) of the spectra become closer to a rectangle, but the sidelobes (two at $N = 4$ and four at $N = 6$) arise. These sidelobes emerge due to the appearance of additional zeros of the reflection and transmission coefficients.

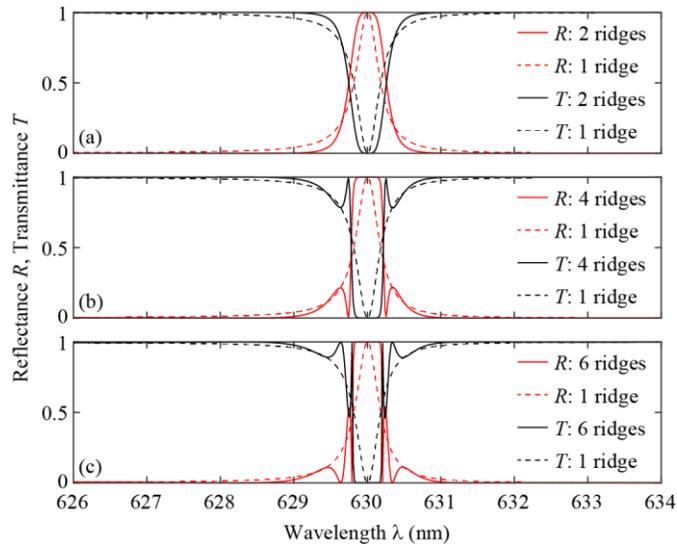

Fig. 4. Transmittance (black) and reflectance (red) spectra of composite structures consisting of $N = 2$ (a), $N = 4$ (b), and $N = 6$ (c) ridges at the ridge width $w = 380$ nm and the distance between the ridges $l = 948$ nm (solid lines). Dashed lines show the spectra of a single ridge.

The shape of the spectra can be altered by changing the widths $l_j$ of the phase-shift regions separating the ridges. Indeed, in the general case, the zeros and poles in Eq. (15) describing the reflection and transmission coefficients of a composite structure comprising $N$ ridges are functions of the variables $l_j$, $j = 1,...,N-1$. Considering $l_j$ as optimization parameters, one can try to obtain the reflectance and transmittance spectra with a required shape. Let us define the desired shape of the reflectance spectrum $R_N(\omega) = |r_N(\omega)|^2$ by the function

$$R_{BW,N}(\lambda) = \frac{1}{1+\left[(\lambda-\lambda_0)/\sigma\right]^{2N}}, \quad (16)$$

which describes a peak with a smoothed rectangular shape having the width $2\sigma$ and no sidelobes. Note that Eq. (16) generalizes Eq. (11) and corresponds to the squared modulus of the transfer function of the Butterworth filter of the order $N$ [24]. The function defined by Eq. (16) can be referred to as Butterworth line shape, which becomes the conventional Lorentzian line shape at $N=1$.

Let us demonstrate the possibility of tailoring the shape of the resonant reflectance peak (transmittance dip) of the composite structures at $N=4$ and $N=6$. The distances $l_j$ were optimized using the conjugate gradient method in order to obtain reflectance spectra described by the functions $R_{BW,4}(\lambda)$ (at $N=4$) and $R_{BW,6}(\lambda)$ (at $N=6$) at $2\sigma = 1$ nm. The used $\sigma$ value corresponds to the reflectance peak width of a single ridge at the 0.1 level (Fig. 3). As a result of the optimization, the following widths of the phase-shift regions were found: $l_1 = l_3 = 1278$ nm and $l_2 = 1395$ nm at $N=4$ and $l_1 = l_5 = 1486$ nm, $l_2 = l_4 = 1383$ nm, and $l_3 = 1485$ nm at $N=6$. The reflectance and transmittance spectra of the composite structures calculated at the obtained values of $l_j$ are shown in Fig. 5. The calculated spectra are very close to the functions $T_{BW,N}(\lambda)$ and $R_{BW,N}(\lambda)$ both at $N=4$ [Fig. 5(a)] and at $N=6$ [Fig. 5(b)].

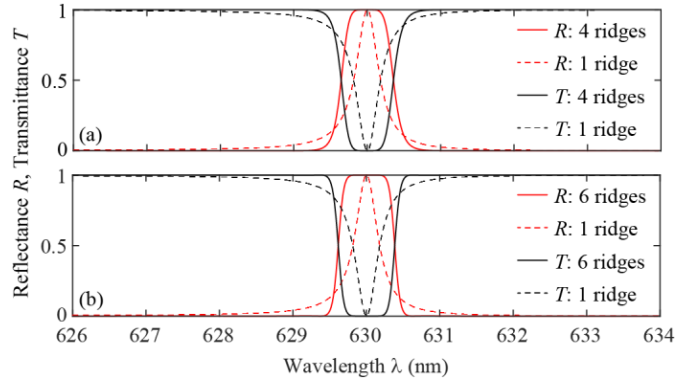

Fig. 5. Transmittance (black) and reflectance (red) spectra of the optimized composite structures consisting of $N=4$ (a) and $N=6$ (b) ridges at $w=380$ nm (solid lines). Dashed lines show the spectra of a single ridge.

Let us now consider an important advantage of the proposed composite filters operating in the near-BIC regime, which consists in the possibility of generating a near-rectangular reflectance peaks with an essentially subnanometer width. The width of the resonance of a single ridge decreases when approaching a BIC and finally vanishes (Fig. 3). Therefore, in the vicinity of a BIC, it is theoretically possible to obtain a Lorentzian reflectance peak with an arbitrarily small width. Let us remind that the width of the flat-top peak of the composite structure is very close to the width of the initial Lorentzian peak (at the 0.1 level). That means that by tuning the width of the ridges constituting the composite structure, one can engineer a very narrow peak having a flat-top (rectangular) shape described by Eq. (16). Let us demonstrate this possibility. In the example considered above ($\theta = \theta_3 = 53.72°$, $w = 380$ nm), the width of the initial Lorentzian-shape reflectance peak at the 0.1 level amounts to $\Delta \approx 1$ nm (Figs. 3 and 5). At the angle of incidence $\theta = \theta_2 = 53.37°$ and the ridge width $w = 360$ nm, the width of the peak (Fig. 3) decreases by approximately 5 times to $\Delta \approx 0.2$ nm. Figure 6 shows the spectra of the composite structures constructed using this "high-Q" ridge with the width $w = 360$ nm and consisting of $N = 2$, $N = 4$, and $N = 6$ ridges. In the structure with $N = 2$, the width of the phase-shift region between the ridges $l = 948$ nm satisfies the condition of Eq. (10). In this case, the reflectance spectrum is very close to the squared modulus of the second-order Butterworth filter. In the case of $N = 4$ and $N = 6$, the widths $l_j$ of the phase-shift regions were optimized in order to obtain the spectra with the Butterworth line shapes described by the functions $R_{BW,4}(\lambda)$ and $R_{BW,6}(\lambda)$, respectively. The following $l_j$ values were obtained: $l_1 = l_3 = 1285$ nm and $l_2 = 1410$ nm at $N = 4$ and $l_1 = l_5 = 1286$ nm, $l_2 = l_4 = 1405$ nm, and $l_3 = 1282$ nm at $N = 6$. Figure 6 shows that the obtained spectra indeed contain smoothed rectangular reflectance peaks with the width of about $0.2$ nm, i.e. approximately 5 times narrower than in Fig. 5 (note the different wavelength ranges shown in Figs. 5 and 6). By further approaching the BIC located at $w = 344$ nm, $\theta = 53.06°$, one can further decrease the width of the resonant peak provided by the resonant structure. It is important to note that, in contrast to many conventional structures providing narrow reflectance or transmittance peaks [1, 7, 8, 11], a decrease in the peak width does not lead to an increase in the structure size.

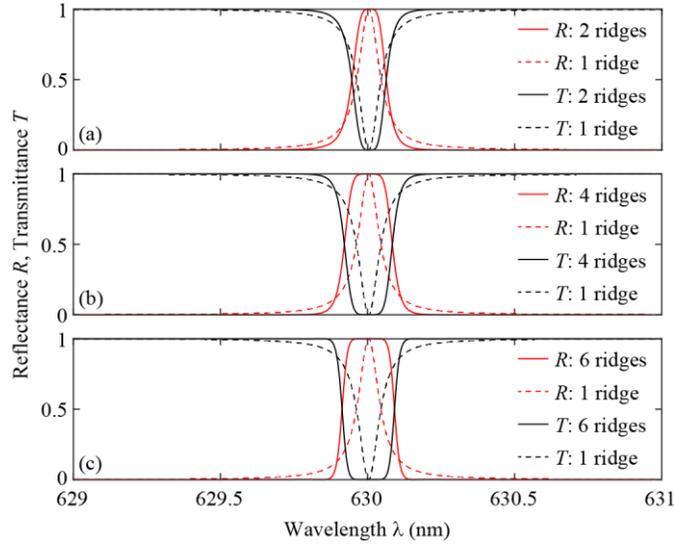

Fig. 6. Transmittance (black) and reflectance (red) spectra of the optimized composite structures consisting of $N = 2$ (a), $N = 4$ (b), and $N = 6$ (c) ridges at $w = 360$ nm (solid lines). Dashed lines show the spectra of a single ridge.

## 5. Fabry–Pérot bound states in the continuum in the composite structure

In addition to the possibility of obtaining a nearly rectangular reflectance peak with flat top, steep slopes and virtually no sidebands, the studied composite structures possess one more remarkable property. Before discussing it, let us remind that the single ridge on the surface of the waveguide used as the building block of the proposed composite structures supports BIC, which is caused by the interaction of TE- and TM-polarized modes in the ridge region [13]. Surprisingly, although the width of the ridges constituting the considered composite filters is detuned from the BIC condition, the composite structure can by itself support BICs, albeit of a different type. Indeed, let us show that if the distances between the ridges satisfy the Fabry–Pérot resonance condition

$$\psi(l_i) + \varphi = \pi m, \; m \in \mathbb{N}, \qquad (17)$$

the so-called Fabry–Pérot bound states in the continuum are formed in the composite structure [25–27]. Let us remind that in Eq. (17), $\psi(l_i)$ is the phase shift acquired by the mode upon propagation through the phase-shift region and defined by Eq. (5), and $\varphi$ is the phase of the reflection coefficient $r_1(\lambda_0)$ in Eq. (2).

First, let us consider the structure comprising two ridges. The reflection and transmission coefficients of this structure are described by Eqs. (7)–(9). If the condition of Eq. (17) is met, one of the poles in Eq. (9) becomes real and coincides with the real-valued transmittance zero. To be specific, let $\operatorname{Im}\lambda_{p,1} = 0$. In this case, the integer $m$ in Eq. (17) is even, and $\lambda_{p,2} = \operatorname{Re}\lambda_p + 2\mathrm{i}\operatorname{Im}\lambda_p$. The eigenmode of the composite structure with the wavelength $\lambda = \operatorname{Re}\lambda_p$ is a BIC [25–27]. The considered structure consisting of two ridges has only two scattering channels corresponding to the reflected and transmitted TE-polarized modes. In the case of a BIC, the leakage to these channels is prevented due to the Fabry–Pérot resonance formed between the ridges under the condition of Eq. (17).

Next, let us consider a composite structure comprising three ridges. The spectra of this structure are described by Eqs. (12)–(14). Under the condition of Eq. (17), the poles defined by Eq. (14) (the complex wavelengths of the eigenmodes of the structure) take the form

$$\lambda_{p,1} = \lambda_{p,2} = \operatorname{Re}\lambda_p, \ \lambda_{p,3} = \operatorname{Re}\lambda_p + 3\mathrm{i}\operatorname{Im}\lambda_p. \tag{18}$$

Therefore, if the condition of Eq. (17) is met, the poles $\lambda_{p,1}$ and $\lambda_{p,2}$ become real and coincide with the real-valued zeros of the reflection and transmission coefficient. This means that two equal-wavelength BICs (or, in other words, a double-degenerate BIC) are formed in the composite structure containing three ridges.

In a composite structure consisting of $N$ ridges, the reflection and transmission coefficients defined by Eq. (15) have $N$ poles. By induction, it can be shown that, under the condition of Eq. (17), $N-1$ poles become real, coinciding with the real-valued zeros of the reflection and transmission coefficients. In this case, ($N-1$)-degenerate BICs are formed in the composite structure. One can show that the only remaining pole will have an imaginary part $N$ times greater than the imaginary part of the pole $\lambda_p$ of the reflection and transmission coefficients of the single ridge: $\lambda_{p,N} = \operatorname{Re}\lambda_p + N\mathrm{i}\operatorname{Im}\lambda_p$. Let us note that the field of the ($N-1$)-degenerate BICs is localized between each pair of adjacent ridges, as opposed to the $N$-degenerate BICs discussed in Section 3, the field of which is localized *inside* each ridge.

Let us note that the formation of the Fabry–Pérot BICs in a composite structure consisting of two resonators is well-known [25–27]. At the same time, to the best of our knowledge, the formation of ($N-1$)-degenerate BICs in a composite structure consisting of $N$ resonant ridges (or other resonators) has not been previously studied.

In order to confirm the formation of the Fabry–Pérot BICs, we calculated the reflectance $R_N(l,\lambda) = |r_N(l,\lambda)|^2$ and the transmittance $T_N(l,\lambda) = |t_N(l,\lambda)|^2$ of several composite structures vs. the wavelength and the distance $l$ between the ridges. The calculation was carried out at the fixed angle of incidence $\theta = 53.72°$ and the ridge width $w = 380\,\mathrm{nm}$. Figure 7 shows the reflectance and transmittance of the composite structures consisting of two, three, and four ridges. Vertical dashed lines show the width of the phase-shift regions $l = l_{FP} = 970.2\,\mathrm{nm}$, which satisfies the Fabry–Pérot resonance condition of Eq. (17) at $m=4$. Horizontal dashed lines show the resonant wavelength of the single ridge $\lambda_0 = 630\,\mathrm{nm}$. In the vicinity of the intersection of these lines, resonant features (reflectance minima and transmittance maxima) are clearly visible. When approaching the intersection, the width of the resonant features decreases (the quality factor of the resonances increases). It is important to note that $N-1$ sharp transmittance peaks occur against the background of a relatively wide

transmittance dip caused by the multiple zero of the transmission coefficient $t_N(\lambda)$ defined by Eq. (15). This effect is similar to the electromagnetically induced transparency (EIT) effect and, in the considered case, is associated with the presence of $N-1$ complex-valued zeros in the reflection coefficient $r_N(\lambda)$. At the vertical dashed lines, the resonant features vanish, which confirms the formation of the BICs. Let us note that according to the presented theoretical description, the number of the vanishing resonant features (and the number of the emerging EIT-peaks) is one less than the number of the ridges $N$, which constitute the composite structure.

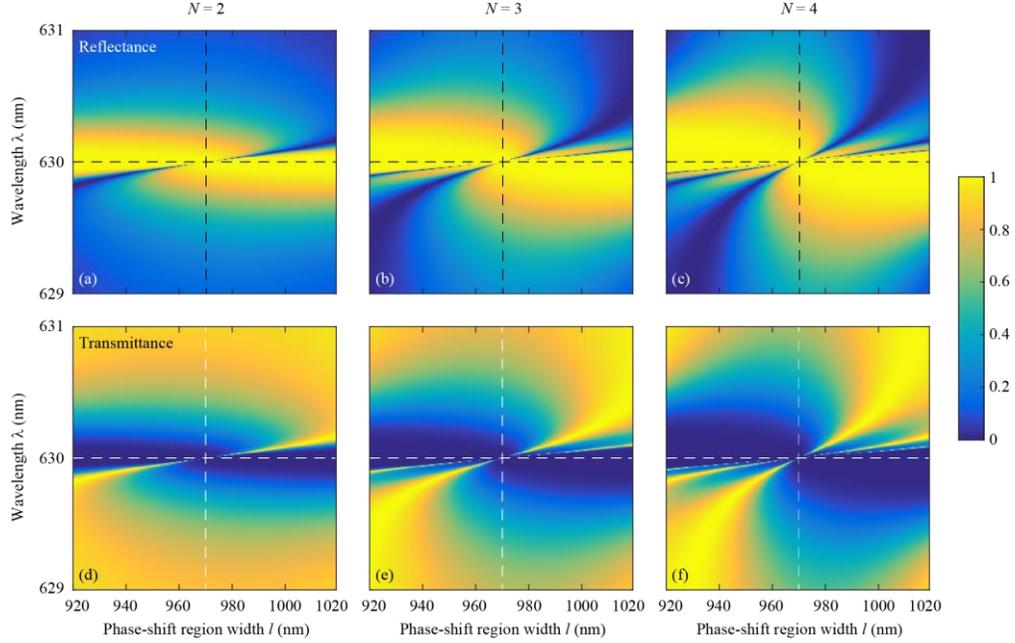

Fig. 7. Reflectance $R_N(l,\lambda)=|r_N(l,\lambda)|^2$ (top) and transmittance $T_N(l,\lambda)=|t_N(l,\lambda)|^2$ (bottom) of the composite structures consisting of $N=2$ (a), (d), $N=3$ (b), (e) and $N=4$ (c), (f) ridges vs. the distance between the ridges $l$ and the free-space wavelength. Vertical dashed lines show the distance $l_{FP}=970.2$ nm corresponding to the Fabry–Pérot resonance.

As an additional illustration of the BIC formation and the EIT effect, let us consider the transmittance spectra of a composite structure consisting of three ridges (Fig. 8). The spectra were calculated at the phase-shift region width $l=l_{FP}=970.2$ nm corresponding to the BIC condition of Eq. (17), and at the "offset" widths $l=l_{FP}+5$ nm $=975.2$ nm and $l=l_{FP}+10$ nm $=980.2$ nm. At $l=l_{FP}$, the transmittance spectrum has a smooth dip centered at $\lambda=\lambda_0=630$ nm. This agrees with Eq. (18), according to which the transmission coefficient has a single pole $\lambda_{p,3}=\operatorname{Re}\lambda_p+3\mathrm{i}\operatorname{Im}\lambda_p$. If we violate the BIC condition by increasing the distance between the ridges by e.g. $\delta=5$ nm, two sharp peaks occur in the transmittance spectra at the wavelengths greater than the central wavelength $\lambda_0$ (see the red curve in Fig. 8). These peaks correspond to the EIT effect and are associated with the zeros of the reflection coefficient defined by Eq. (13). At $\delta=10$ nm, the EIT-peaks become wider and experience a further red shift (see the yellow curve in Fig. 8). It is interesting to note that the different widths of the EIT-peaks follow from the expression for the poles of the transmission coefficient [Eq. (14)]. Indeed, the widths of the EIT-peaks are determined by the magnitudes of the imaginary part of the poles $\lambda_{p,i}, i = 1, 2$. From Eq. (14), it is easy to obtain that near the BIC [in the vicinity of the point $l = l_{FP}$ satisfying the condition of Eq. (17)], the imaginary parts of the poles $\lambda_{p,i}$ are quadratic with respect to the "offset" magnitude $\delta$:

$$\operatorname{Im} \lambda_{p,i}(\delta) = \alpha_i \delta^2 + O(\delta^3), \ i = 1, 2, \qquad (19)$$

where $\alpha_1 = 2$ and $\alpha_2 = 2/27$. Since $\alpha_1/\alpha_2 = 27$, the EIT-peak corresponding to the pole $\lambda_{p,2}$ should have an order-of-magnitude smaller width, which agrees with the spectra shown in Figs. 7(e) and 8.

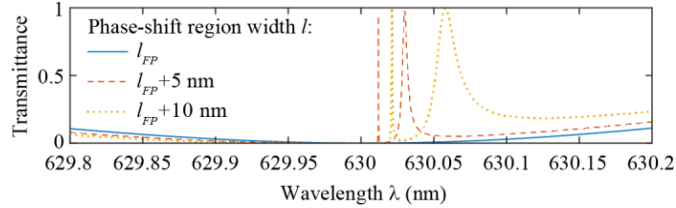

Fig. 8. Transmittance spectra of the composite structure consisting of $N = 3$ ridges calculated at the width of the phase-shift region between the ridges $l = l_{FP} = 970.2$ nm corresponding to the BIC condition (solid blue line), and at the widths $l = l_{FP} + 5$ nm $= 975.2$ nm (dashed red line) and $l = l_{FP} + 10$ nm $= 980.2$ nm (dotted yellow line).

## 6. Conclusion

In the present work, we investigated resonant optical properties of composite structures consisting of $N$ identical ridges on the surface of a slab waveguide separated by phase-shift regions. The single ridge used as the building block of the proposed composite structures supports bound states in the continuum and high-Q resonances having a Lorentzian line shape. Using the scattering matrix formalism, we obtained resonant approximations of the reflection and transmission coefficients of the composite structure. According to the derived expressions, in the case when the ridges constituting the composite structure support BICs, the composite structure supports $N$-degenerate BICs, whereas the detuning from the BIC condition allows one to obtain flat-top reflection filters. In particular, we demonstrated that at $N = 2$, the reflection coefficient coincides with the transfer function of the second-order Butterworth filter. It was shown that by optimizing the widths of the phase-shift regions at $N > 2$, one can obtain a flat-top reflectance peak with steep slopes using just a few ridges. In our opinion, the proposed composite structure has a simpler geometry and is more compact than the planar filters based on integrated Bragg gratings.

To the best of our knowledge, this work presents the first theoretical and numerical demonstration of integrated flat-top filters operating in the near-BIC regime. An important advantage of the proposed structure consists in the fact that the flat-top reflectance peak can be made arbitrarily narrow by choosing the width of the ridges in the vicinity of a BIC supported by a single ridge. In this case, the corresponding composite structure can be considered as a perturbation of the composite structure supporting an $N$-degenerate BIC. In contrast to many conventional structures providing narrow reflectance or transmittance peaks, a decrease in the peak width does not lead to an increase in the structure size.

In addition to the spectral filtering applications, the considered composite structures possess one more remarkable optical property, namely, the formation of bound states in the continuum having a different origin than that of the BICs supported by the initial ridge. At the distances between the adjacent ridges satisfying the Fabry–Pérot condition, $(N-1)$-degenerate Fabry–Pérot BICs arise in the composite structure comprising $N$ ridges. In the vicinity of the formed BICs, an effect very similar to the electromagnetically induced transparency (EIT) effect is observed. In the present case, the EIT peaks can be explained by the presence of $N-1$ zeros of the reflection coefficient.

The obtained results may find application in the design of optical filters, sensors, and the devices for the transformation of optical signals. In particular, the proposed composite structures can be used for higher-order differentiation of optical signals propagating in the slab waveguide. In addition, the theoretical result presented in Section 3 and concerning the description of the resonant properties of the composite structures are not specific for the considered ridge structures and can be extended to other composite structures comprising Lorentzian-line-shape resonators separated by phase-shift layers.

## Funding

Russian Foundation for Basic Research (18-37-20038); Russian Federation Ministry of Science and Higher Education (State contract with the "Crystallography and Photonics" Research Centre of the RAS under agreement 007-GZ/Ch3363/26); Russian Science Foundation (19-19-00514).

## Acknowledgments

This work was funded by Russian Foundation for Basic Research (design and investigation of the composite filters, Sections 3 and 4), by the Russian Federation Ministry of Science and Higher Education (simulation of the single-ridge structure, Section 2), and by Russian Science Foundation (investigation of bound states in the continuum and electromagnetically induced transparency in the composite structures, Section 5).

## References

1. H. A. Haus, *Waves and Fields in Optoelectronics* (Prentice-Hall, 1984).
2. T. Mossberg, "Planar holographic optical processing devices," Opt. Lett. **26**(7), 414–416 (2001).
3. G. Calafiore, A. Koshelev, S. Dhuey, A. Goltsov, P. Sasorov, S. Babin, V. Yankov, S. Cabrini, and C. Peroz, "Holographic planar lightwave circuit for on-chip spectroscopy," Light: Sci. Appl. **3**(9), e203 (2014).
4. S. Babin, A. Bugrov, S. Cabrini, S. Dhuey, A. Goltsov, I. Ivonin, E.-B. Kley, C. Peroz, H. Schmidt, and V. Yankov, "Digital optical spectrometer-on-chip," Appl. Phys. Lett. **95**(4), 041105 (2009).
5. C. Peroz, C. Calo, A. Goltsov, S. Dhuey, A. Koshelev, P. Sasorov, I. Ivonin, S. Babin, S. Cabrini, and V. Yankov, "Multiband wavelength demultiplexer based on digital planar holography for on-chip spectroscopy applications," Opt. Lett. **37**(4), 695–697 (2012).
6. C. Peroz, A. Goltsov, S. Dhuey, P. Sasorov, B. Harteneck, I. Ivonin, S. Kopyatev, S. Cabrini, S. Babin, and V. Yankov, "High-resolution spectrometer-on-chip based on digital planar holography," IEEE Photon. J. **3**(5), 888–896 (2011).
7. L. L. Doskolovich, E. A. Bezus, and D. A. Bykov, "Two-groove narrowband transmission filter integrated into a slab waveguide," Photon. Res. **6**(1), 61–65 (2018).
8. R. V. Schmidt, D. C. Flanders, C. V. Shank, and R. D. Standley, "Narrow-band grating filters for thin-film optical waveguides," Appl. Phys. Lett. **25**(11), 651–652 (1974).
9. C. S. Hong, J. B. Shellan, A. C. Livanos, A. Yariv, and A. Katzir, "Broad-band grating filters for thin film optical waveguides," Appl. Phys. Lett. **31**(4), 276–278 (1977).
10. L. A. Weller-Brophy and D. G. Hall, "Analysis of waveguide gratings: application of Rouard's method," J. Opt. Soc. Am. A **2**(6), 863–871 (1985).
11. R. Zengerle and O. Leminger, "Phase-shifted Bragg grating filters with improved transmission characteristics," J. Lightwave Technol. **13**(12), 2354–2358 (1995).
12. J. N. Damask and H. A. Haus, "Wavelength-division multiplexing using channel-dropping filters," J. Lightwave Technol. **11**(3), 424–428 (1993).
13. E. A. Bezus, D. A. Bykov, and L. L. Doskolovich, "Bound states in the continuum and high-Q resonances supported by a dielectric ridge on a slab waveguide," Photon. Res. **6**(11), 1084–1093 (2018).
14. E. A. Bezus, L. L. Doskolovich, D. A. Bykov, and V. A. Soifer, "Spatial integration and differentiation of optical beams in a slab waveguide by a dielectric ridge supporting high-Q resonances," Opt. Express **26**(19), 25156–25165 (2018).
15. C.-L. Zou, J.-M. Cui, F.-W. Sun, X. Xiong, X.-B. Zou, Z.-F. Han, and G.-C. Guo, "Guiding light through optical bound states in the continuum for ultrahigh-Q microresonators," Laser Photon. Rev. **9**(1), 114–119 (2015).
16. A. P. Hope, T. G. Nguyen, A. Mitchell, and W. Bogaerts, "Quantitative analysis of TM lateral leakage in foundry fabricated silicon rib waveguides," IEEE Photon. Technol. Lett. **28**(4), 493–496 (2016).

17. Y. H. Ko and R. Magnusson, "Flat-top bandpass filters enabled by cascaded resonant gratings," Opt. Lett. **41**(20), 4704–4707 (2016).
18. E. Silberstein, P. Lalanne, J.-P. Hugonin, and Q. Cao, "Use of grating theories in integrated optics," J. Opt. Soc. Am. A **18**(11), 2865–2875 (2001).
19. J. P. Hugonin and P. Lalanne, "Perfectly matched layers as nonlinear coordinate transforms: a generalized formalization," J. Opt. Soc. Am. A **22**(9), 1844–1849 (2005).
20. L. Li, "Formulation and comparison of two recursive matrix algorithms for modeling layered diffraction gratings," J. Opt. Soc. Am. A **13**(5), 1024–1035 (1996).
21. N. A. Gippius, S. G. Tikhodeev, and T. Ishihara, "Optical properties of photonic crystal slabs with an asymmetrical unit cell," Phys. Rev. B **72**(4), 045138 (2005).
22. D. A. Bykov, L. L. Doskolovich, N. V. Golovastikov, and V. A. Soifer, "Time-domain differentiation of optical pulses in reflection and in transmission using the same resonant grating," J. Opt. **15**(10), 105703 (2013).
23. N. V. Golovastikov, D. A. Bykov, and L. L. Doskolovich, "Temporal differentiation and integration of 3D optical pulses using phase-shifted Bragg gratings," Comput. Opt. **41**(1), 13–21 (2017).
24. H.-C. Liu and A. Yariv, "Synthesis of high-order bandpass filters based on coupled-resonator optical waveguides (CROWs)," Opt. Express **19**(18), 17653–17668 (2011).
25. M. F. Limonov, M. V. Rybin, A. N. Poddubny, and Y. S. Kivshar, "Fano resonances in photonics," Nat. Photon. **11**(9), 543–554 (2017).
26. D. C. Marinica, A. G. Borisov, and S. V. Shabanov, "Bound states in the continuum in photonics," Phys. Rev. Lett. **100**(18), 183902 (2008).
27. V. Liu, M. Povinelli, and S. Fan, "Resonance-enhanced optical forces between coupled photonic crystal slabs," Opt. Express **17**(24), 21897–21909 (2009).